\documentclass[apj]{emulateapj}
\usepackage{mathptmx}

\newcommand{\ltsim}{\raisebox{-.5ex}{$\;\stackrel{<}{\sim}\;$}}
\newcommand{\gtsim}{\raisebox{-.5ex}{$\;\stackrel{>}{\sim}\;$}}
\newcommand{\kms}{\ifmmode {\rm km\ s}^{-1} \else km s$^{-1}$\fi}
\newcommand{\vFWHM}{\ifmmode V_{\mbox{\tiny FWHM}} \else
            $V_{\mbox{\tiny FWHM}}$\fi}
\newcommand{\msun}{$M_{\odot}$}
\newcommand{\et}{et al.\ }

\newcommand{\mbh}{$M_{\rm BH}$}
\newcommand{\aox}{$\alpha_{\rm ox}$}
\newcommand{\nh}{$N_{\rm H}$}
\newcommand{\Ka}{\hbox{Fe K$\alpha$}}
\newcommand{\xmm}{{\hbox{\sl XMM-Newton}}}
\newcommand{\chandra}{{\sl Chandra}}

\journalinfo{The Astronomical Journal, 
???:??--??, 200? ???, astro-ph/0509146}
\slugcomment{Received 2005 July 11; accepted 2005 September 4}

\shorttitle{{\sl XMM-NEWTON} SPECTROSCOPY OF CSO~755}
\shortauthors{SHEMMER ET AL.}

\begin{document}
\title{{\sl XMM-Newton} Spectroscopy of the Highly Polarized and
  Luminous Broad Absorption Line Quasar CSO~755}
\author{
O.~Shemmer,\altaffilmark{1}
W.~N.~Brandt,\altaffilmark{1}
S.~C.~Gallagher,\altaffilmark{2}
C.~Vignali,\altaffilmark{3,4}
Th.~Boller,\altaffilmark{5}
G.~Chartas,\altaffilmark{1}
and A.~Comastri\altaffilmark{4}
}

\altaffiltext{1} {Department of Astronomy \& Astrophysics, The
  Pennsylvania State University, University Park, PA 16802, USA;
  ohad@astro.psu.edu}

\altaffiltext{2} {Department of Physics \& Astronomy, University of
  California -- Los Angeles, Mail Code 154705, 475 Portola Plaza, Los
  Angeles CA, 90095--4705, USA}

\altaffiltext{3} {Dipartimento di Astronomia, Universita` degli Studi
  di Bologna, Via Ranzani 1, 40127 Bologna, Italy}

\altaffiltext{4} {INAF - Osservatorio Astronomico di Bologna, Via
  Ranzani 1, 40127 Bologna, Italy}

\altaffiltext{5} {Max-Planck-Institut f\"{u}r extraterrestrische
  Physik, Postfach 1312, 85741 Garching, Germany}

\begin{abstract}
We present the results from \xmm\ observations of the highly optically
polarized broad absorption line quasar (BALQSO) CSO~755. By analyzing
its \hbox{X-ray} spectrum with a total of $\sim$3000 photons we find
that this source has an \hbox{X-ray} continuum of `typical'
radio-quiet quasars, with a photon index of
$\Gamma$=1.83$^{+0.07}_{-0.06}$, and a rather flat (\hbox{X-ray}
bright) intrinsic optical-to-X-ray spectral slope of \aox=$-$1.51. The
source shows evidence for intrinsic absorption, and fitting the
spectrum with a neutral-absorption model gives a column density of
\nh$\sim$1.2$\times$10$^{22}$~cm$^{-2}$; this is among the lowest
\hbox{X-ray} columns measured for BALQSOs.  We do not detect, with
high significance, any other absorption features in the X-ray
spectrum.  Upper limits we place on the rest-frame equivalent width of
a neutral (ionized) \Ka\ line, $\leq$180~eV ($\leq$120~eV), and on the
Compton-reflection component parameter, $R\leq$0.2, suggest that most
of the \hbox{X-rays} from the source are directly observed rather than
being scattered or reflected; this is also supported by the relatively
flat intrinsic \aox\ we measure.  The possibility that most of the
\hbox{X-ray} flux is scattered due to the high level of
\hbox{UV--optical} polarization is ruled out.  Considering data for 46
BALQSOs from the literature, including CSO~755, we have found that the
\hbox{UV--optical} continuum polarization level of BALQSOs is not
correlated with any of their \hbox{X-ray} properties.  A lack of
significant short- and long-term \hbox{X-ray} flux variations in the
source may be attributed to a large black-hole mass in CSO~755.  We
note that another luminous BALQSO, PG~2112$+$059, has both similar
shallow \ion{C}{4} BALs and moderate \hbox{X-ray} absorption.
\end{abstract}


\section{Introduction}
\label{introduction}

Broad absorption line quasars (BALQSOs) are well known to be weak
X-ray sources, and their \hbox{X-ray} fluxes cannot be predicted well
from their optical brightnesses (e.g., Green \& Mathur 1996; Brinkmann
\et 1999; Gallagher \et 1999; Brandt, Laor, \& Wills 2000; Green \et
2001; Strateva \et 2005).  High-quality \xmm\ and \chandra\ spectra of
$\approx$10 BALQSOs obtained during the past five years have shown
that their intrinsic column densities are typically
$\approx$10$^{23}$~cm$^{-2}$ (e.g., Gallagher \et 2002; Chartas
\et~2002, 2003; Grupe \et 2003). In at least two of these cases, such
X-ray spectra have also revealed discrete absorption features
suggesting the presence of relativistic outflows (Chartas \et 2002,
2003); however, the physical connection between the characteristic
ultraviolet (UV) and \hbox{X-ray} absorption is not yet clear.

BALQSOs are known to have higher \hbox{UV--optical} polarization on
average compared with `typical' quasars, and $\approx$20\% of BALQSOs
have a continuum polarization level of $\geq$2\% (e.g., Hutsem{\'
  e}kers, Lamy, \& Remy 1998; Schmidt \& Hines 1999). The polarization
probably arises due to light scattered off an electron or dust
`mirror' of moderate optical depth near the center (e.g., Schmidt \&
Hines 1999), and it is strongly anti-correlated with the BALQSO
detachment index (DI; Weymann \et 1991), i.e., sources with UV
absorption profiles with lower onset velocities show a higher level of
\hbox{UV--optical} continuum polarization (e.g., Ogle \et 1999; Lamy
\& Hutsem{\' e}kers 2004). Measurements of the polarization levels of
the continuum, broad emission lines, and BAL troughs provide a
valuable probe for tracing the inner structure in these sources. Based
upon \hbox{X-ray} ({\sl ASCA}) observations of a sample of seven
BALQSOs, Gallagher \et (1999) tentatively suggested that highly
polarized BALQSOs (with a level of \hbox{2.3\%--5\%} of broad-band
optical continuum polarization) may also be the \hbox{X-ray} brightest
of that class (see also Brandt \et 1999).  Such a trend could appear,
for example, if most of the \hbox{X-ray} light is blocked from direct
view by thick material ($\gtsim$10$^{24}$~cm$^{-2}$) associated with
the BAL flow, and the \hbox{X-rays} we observe are mostly due to
scattering. Any connection between \hbox{UV--optical} continuum
polarization and \hbox{X-ray} spectral properties would provide an
important clue about the geometry of matter in BALQSO nuclei.

Case Stellar Object~755 (hereafter CSO~755; Sanduleak \& Pesch 1989;
also known as SBS~1524$+$517 with J2000.0 coordinates:
$\alpha$=15:25:53.9, $\delta$=$+$51:36:49) is a radio-quiet BALQSO at
$z$=2.88.  It is among the most luminous quasars known with
$M_B$=$-$28.4, being the ninth most luminous quasar in the Sloan
Digital Sky Survey (SDSS; York \et 2000) Data Release~3 (Schneider \et
2005). CSO~755 is most probably a high-ionization BALQSO (HiBAL),
since it does not exhibit low-ionization BALs of Al and Fe in the SDSS
spectrum, and it is not heavily reddened.  A comparison between the UV
absorption features in the SDSS spectrum (taken in 2002) and those
observed in earlier spectra of the source (e.g., Korista \et 1993;
Glenn, Schmidt, \& Foltz 1994; Ogle \et 1999) suggests that there have
been no strong variations in the properties of the UV absorber.  The
source is also one of the most optically polarized BALQSOs
($P_{V}\simeq$3.5\%), even though its absorption troughs are more
detached than those of most high polarization BALQSOs (e.g., Glenn \et
1994; Ogle 1998; Ogle \et 1999). Based upon spectropolarimetric
observations of the source, Ogle (1998) was able to constrain the
angular size of the electron-scattering region to be intermediate
between the angular sizes of the high and low velocity BAL clouds in
CSO~755. Combining \hbox{X-ray} data with the available
\hbox{UV--optical} information on the source may provide further
constraints on the geometry and structure of the BAL flow and on the
properties of the scattering medium, since the size of the direct
\hbox{X-ray} continuum source is expected to be much smaller compared
to the sizes of both the UV BALs and the scattering medium responsible
for the UV polarization.

The first sensitive \hbox{X-ray} observations of CSO~755 were obtained
by {\sl BeppoSAX} in 1999 (Brandt \et 1999), and it was also
tentatively detected by the {\sl ROSAT} All Sky Survey. The {\sl
  BeppoSAX} observations provided only loose constraints on the X-ray
spectral shape of the source and a flux measurement which enabled
planning of follow-up \hbox{X-ray} observations. In this paper we
present new, high-quality, \xmm\ observations of CSO~755.  In
\S~\ref{observations} we describe our observations and their
reduction, and in \S~\ref{results} we present the results of the
\hbox{X-ray} spectral analysis and variability of the source. In
\S~\ref{discussion} we discuss our results and the relations between
the UV and \hbox{X-ray} properties of CSO~755 as well as those of
other BALQSOs from the literature. Throughout the~paper we use the
standard cosmological model, with parameters $\Omega_{\Lambda}$=0.7,
$\Omega_{\rm M}$=0.3, and $H_0$=70~\kms~Mpc$^{-1}$ (Spergel \et 2003).

\section{Observations and Data Reduction}
\label{observations}

We obtained imaging spectroscopic observations of CSO~755 with
\xmm\ (Jansen \et 2001) on \hbox{2001 December 8--9} and on
2001~December~13 (hereafter the first and second observations,
respectively). The \xmm\ observation log appears in
Table~\ref{obs_log}.  The data were processed using standard {\sc
  sas\footnote{\xmm\ Science Analysis System.  See
    http://xmm.vilspa.esa.es/external/xmm\_sw\_cal/sas\_frame.shtml}
  v6.1.0} and {\sc ftools} tasks. The event files of both observations
were filtered to include events with {\sc flag}$=$0, and {\sc
  pattern}$\leq$12 ({\sc pattern}$\leq$4) and 200$\leq${\sc
  pi}$\leq$12000 (150$\leq${\sc pi}$\leq$15000) for the MOS (pn)
detectors.  The event files of the first observation were also
filtered to remove a $\sim$7~ks period of flaring activity at the end
of that observation, which was apparent in the light curve of the
entire full-frame window.  The event files of the second observation
were not filtered in time, since more than 90\% of the observation was
performed during an intense background flaring period with background
count rates $\sim$15 times higher than nominal values. The exposure
times listed in Table~\ref{obs_log} reflect the filtered data used in
the analysis.

\begin{deluxetable}{cccc}
\tablecolumns{4}
\tablecaption{Log of \xmm\ Observations of CSO~755
\label{obs_log}}
\tablehead
{
\colhead{Observation} &
\multicolumn{3}{c}{Net Exposure Time (ks) / Source Counts} \\
\colhead{Start Date} &
\colhead{MOS1} &
\colhead{MOS2} &
\colhead{pn}
}
\startdata
2001 December 08 & 29.0 / 604 & 29.0 / 584 & 24.4 / 1920 \\
2001 December 13 & 14.5 / 334 & 14.4 / 310 & 9.7 / 766
\enddata
\end{deluxetable}

To extract \hbox{X-ray} spectra we used a source-extraction aperture
radius of 30\arcsec\ in the images of all three European Photon
Imaging Camera (EPIC) detectors.\footnote{The two Reflection Grating
  Spectrometer detectors on \xmm\ do not have sufficient counts to
  perform a high-resolution spectral analysis for CSO~755.} Background
regions were taken to be at least as large as the source regions;
these were annuli (circles) for the MOS (pn) detectors. The
redistribution matrix files (RMFs; which include information on the
detector gain and energy resolution) and the ancillary response files
(ARFs; which include information on the effective area of the
instrument, filter transmission, and any additional energy-dependent
efficiencies) for the spectra were created with the {\sc sas} tasks
{\sc rmfgen} and {\sc arfgen}, respectively.  The spectra were grouped
with a minimum of 25 counts per bin using the task {\sc grppha}.

\section{X-ray properties of CSO~755}
\label{results}
\subsection{Photon Index and Intrinsic Absorption}
\label{spectral}

\begin{figure}
\plotone{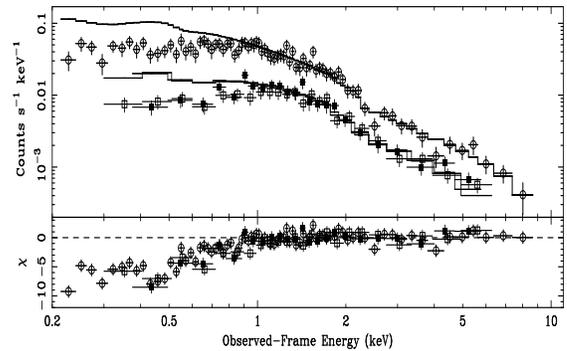}
\caption
{Data, best-fit spectra, and residuals for the first \xmm\ observation
  of CSO~755. Open circles, filled squares, and open squares represent
  the pn, MOS1, and MOS2 data, respectively. Solid lines represent the
  best-fit model for each spectrum, and the thick line marks the
  best-fit model for the pn data.  The data were fitted with a
  Galactic-absorbed power-law in the \hbox{5--30~keV} rest-frame
  energy range (\hbox{$\sim$1.3--7.7}~keV in the observed frame), and
  extrapolated to as low as 0.8~keV in the rest frame. The $\chi$
  residuals are in units of $\sigma$ with error bars of size one.}
\label{figure_abs}
\end{figure}

\begin{figure*}
\plotone{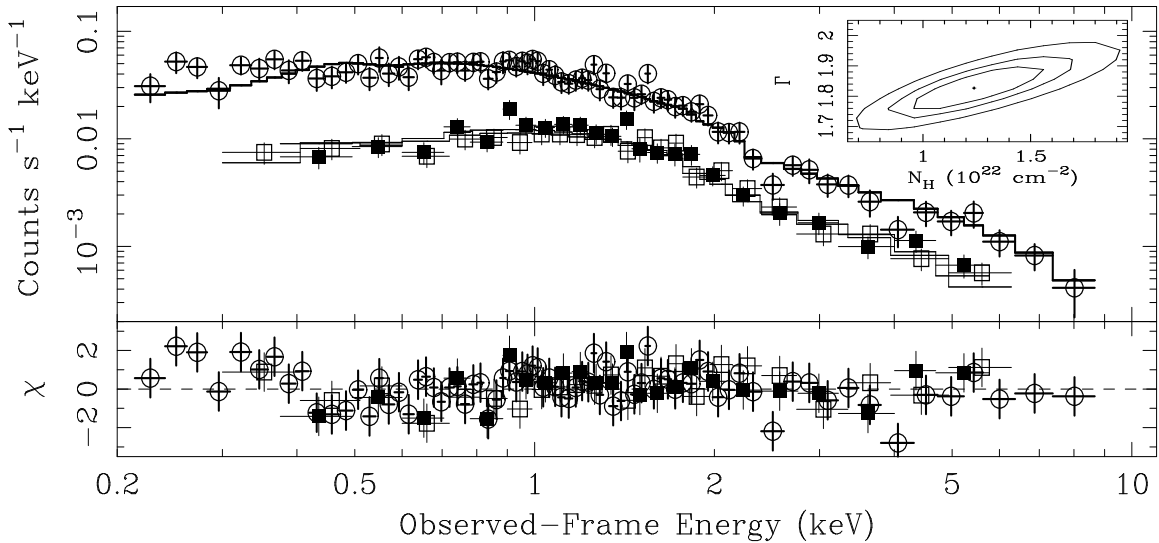}
\caption
{Data, best-fit spectra, and residuals for the first \xmm\ observation
  of CSO~755. Symbols are the same as those of Fig.~\ref{figure_abs}.
  The $\chi$ residuals are in units of $\sigma$ with error bars of
  size one. The inset shows 68, 90, and 99\% confidence contours for
  the intrinsic absorption (\nh) and photon index ($\Gamma$).}
\label{spectrum_figure}
\end{figure*}

We used {\sc xspec v11.3.0} (Arnaud 1996) to fit jointly the data from
all three EPIC detectors in the first observation with a
Galactic-absorbed power law in the rest-frame \hbox{5--30~keV} energy
range (\hbox{$\sim$1.3--7.7}~keV in the observed frame); the Galactic
absorption (Dickey \& Lockman 1990) was fixed to the value
1.57$\times$10$^{20}$~cm$^{-2}$ (obtained using the HEASARC
\nh\
tool\footnote{http://heasarc.gsfc.nasa.gov/cgi-bin/Tools/w3nh/w3nh.pl}).
The fit in the rest-frame \hbox{5--30~keV} band is acceptable with a
derived power-law photon index of $\Gamma$=1.94$\pm$0.12. However,
extrapolation of this model to lower rest-frame energies reveals
strong negative residuals; these indicate the presence of intrinsic
X-ray absorption (Fig.~\ref{figure_abs}).  We therefore added an
intrinsic (redshifted) neutral absorption component with solar
abundances to the model and fitted the data of each EPIC detector
alone, jointly fitted the data of the two MOS detectors, and jointly
fitted all three EPIC detectors.  The best-fit results of this model
for the two observations are summarized in
Table~\ref{0201_0401_fit}. The best-fit spectral parameters for all
three EPIC detectors are consistent within the errors.  The EPIC
spectra from the first observation and their joint, best-fit model
appear in Fig.~\ref{spectrum_figure}, which also includes a
confidence-contour plot of the \hbox{$\Gamma$--\nh} parameter space.

The $\sim$2$\sigma$ residuals apparent in Fig.~\ref{0201_0401_fit} at
observed-frame energies $\ltsim$0.4~keV, may be due to a combination
of calibration uncertainties (e.g., Sembay \et 2004; Kirsch \et 2004)
and intrinsic absorption that is somewhat more complex than the simple
neutral-absorption model we have applied (perhaps due to the partial
covering, ionization, or internal velocity structure of the
absorber). To investigate the nature of these residuals and to check
whether some of the absorption could be due to ionized gas, we fit the
data with a model consisting of a Galactic-absorbed power-law and an
ionized absorber using the {\sc xspec} model {\sc absori} (Done \et
1992; Magdziarz \& Zdziarski 1995). We found a photon index of
1.99$\pm$0.10, and an ionized column density of
\nh=1.0$\pm$0.3$\times$10$^{23}$~cm$^{-2}$.  At face value, an
$F$-test showed that this model provided a better fit to the data over
the model that used a neutral intrinsic absorber
[$\chi^2$(DOF)=85.2(120) for jointly fitting the three EPIC detectors;
  compare with Table~\ref{0201_0401_fit}], and the best-fit spectrum
showed that the $\sim$2$\sigma$ residuals observed at $\ltsim$0.4~keV
have almost disappeared.  However, the use of an ionized absorber
model in our case is almost certainly a simplistic approach to a more
complex problem of BAL flows.  More detailed modeling of AGN outflows
has been performed for several moderate-luminosity sources (e.g.,
Netzer \et 2003; Chelouche \& Netzer 2005), but appropriate models
have not yet been developed for luminous BALQSOs. Moreover, due to the
high redshift of CSO~755, $z$=2.88, any modeling is highly influenced
by the low observed-frame energy bins at the edge of the EPIC response
($\ltsim$0.4~keV). Any fitting carried out in this energy range
depends on a handful of data points, and is somewhat unreliable due to
the calibration uncertainties discussed above.

\begin{deluxetable*}{lcccccc}
\tablecolumns{7}
\tablecaption{Best-Fit \hbox{X-ray} Spectral Parameters for CSO~755.
\label{0201_0401_fit}}
\tablehead
{
\colhead{} &
\multicolumn{3}{c}{2001 December 8} &
\multicolumn{3}{c}{2001 December 13} \\
\colhead{} &
\colhead{} &
\colhead{} &
\colhead{} &
\colhead{} &
\colhead{} &
\colhead{} \\
\colhead{Detector} &
\colhead{\nh \tablenotemark{a}} &
\colhead{$\Gamma$} &
\colhead{$\chi^{2}$(DOF)} &
\colhead{\nh \tablenotemark{a}} &
\colhead{$\Gamma$} &
\colhead{$\chi^{2}$(DOF)} \\
}
\startdata
PN           & $1.17^{+0.37}_{-0.35}$ & $1.86^{+0.09}_{-0.09}$ & $68.7(75)$ &
$1.85^{+2.17}_{-1.41}$ & $1.70^{+0.49}_{-0.18}$ & $135.9(122)$ \\ \\
MOS1         & $2.40^{+1.21}_{-0.99}$ & $1.92^{+0.18}_{-0.17}$ & $16.2(21)$ &
$\le1.43$ & $1.80^{+0.39}_{-0.27}$ & $18.3(31)$ \\ \\
MOS2         & $1.07^{+0.87}_{-0.75}$ & $1.71^{+0.17}_{-0.15}$ & $14.6(21)$ &
$\le1.86$ & $1.71^{+0.44}_{-0.33}$ & $16.5(28)$ \\ \\
MOS1+MOS2    & $1.67^{+0.71}_{-0.63}$ & $1.81^{+0.12}_{-0.12}$ & $33.2(44)$ &
$\le1.20$ & $1.76^{+0.27}_{-0.23}$ & $35.2(61)$ \\ \\
MOS1+MOS2+PN & $1.24^{+0.31}_{-0.31}$ & $1.83^{+0.07}_{-0.06}$ & $107.6(121)$ &
$0.89^{+0.83}_{-0.71}$ & $1.74^{+0.24}_{-0.21}$ & $177.6(185)$ \\
\enddata
\tablecomments{The best-fit intrinsic
  absorption (\nh), photon index ($\Gamma$), normalization, and
  $\chi^2$ were obtained from a model consisting of a power law with
  both Galactic and neutral intrinsic absorption. The lack of
  low-energy response and low signal-to-noise ratio make MOS detector
  data less sensitive for constraining \nh\ in the second
  observation.} 
\tablenotetext{a}{Intrinsic column density in units
  of 10$^{22}$~cm$^{-2}$. Errors are calculated taking one parameter
  to be of interest ($\Delta \chi^{2}=2.71$; e.g., Avni 1976).}
\end{deluxetable*}

When fit with a neutral-absorption model, CSO~755 has a relatively low
\hbox{X-ray} column density of
\nh=1.2$\pm$0.3$\times$10$^{22}$~cm$^{-2}$ (Table~\ref{0201_0401_fit}
and Fig.~\ref{spectrum_figure}), compared to other BALQSOs
(\nh$\sim$10$^{23}$~cm$^{-2}$) fit with the same model (e.g.,
Gallagher \et 2002). The power-law photon index in the rest-frame
\hbox{$\sim$0.8--30~keV} band, $\Gamma$=1.83$^{+0.07}_{-0.06}$, is
well within the typical range of indices for other BALQSOs (Gallagher
\et 2002) and, in fact, is within the range for all radio-quiet type~1
(i.e., unobscured) AGNs (e.g., Reeves \& Turner 2000; Shemmer \et
2005; Vignali \et 2005). This supports the conclusion that intrinsic
BALQSO \hbox{X-ray} continua are not different from those of `typical'
radio-quiet type~1 AGNs (e.g., Gallagher \et 2002).

The analyses performed throughout the rest of this paper are based
upon the data of all three EPIC detectors from the first observation.
The data from the second observation, which were not filtered in time,
exhibit poor signal-to-noise ratios which result in large
uncertainties on the best-fit parameters (although these results are
generally in agreement with those based upon the data from the first
observation; see Table~\ref{0201_0401_fit}).

\subsection{Iron Features and Compton Reflection}
\label{Ka}

The EPIC data show a hint of the presence of an iron absorption edge
in the rest-frame energy range \hbox{7.1--9.3~keV}, where the
low-energy end of this range corresponds to the threshold energy of
\ion{Fe}{1}, and the high-energy end corresponds to the threshold
energy of \ion{Fe}{26} (Verner \et 1993). In order to test for the
existence of an iron edge we fit a model consisting of a
Galactic-absorbed power law, a neutral intrinsic absorption component,
and an iron-edge component (the EPIC spectral resolution should, in
principle, allow us to resolve such an edge).  We fitted the data with
this model first using a neutral edge, and then fitted an ionized
edge. The threshold of the neutral edge was fixed at a rest-frame
energy of 7.1~keV, and the threshold of the ionized edge was fixed at
rest-frame 9.3~keV; the strength of the edge was free to vary.  Using
an \hbox{$F$-test} we found that the model using a neutral iron edge
did not significantly improve the fit over the model we used in
\S~\ref{spectral}, and the upper limit we obtained on the absorption
depth of a neutral edge is $\tau\leq$0.1. Fitting the data with an
ionized iron edge significantly improves the fit from
\S~\ref{spectral}; we obtained a $\chi^2$(DOF)=102.1(120), compared
with the $\chi^2$(DOF)=107.6(121) obtained using the model of
\S~\ref{spectral} (see also Table~\ref{0201_0401_fit}). The absorption
depth of an ionized edge is $\tau$=0.4$^{+0.2}_{-0.3}$. However, the
suggested ionized iron edge, at an observed energy of $\sim$2.4~keV,
coincides with the pronounced Au~M edge in the EPIC spectral response,
leading to some uncertainty about its existence.

Some BALQSOs have shown evidence for \hbox{X-ray} `reflection' via
either \Ka\ lines or Compton-reflection `humps' (e.g., Oshima \et
2001; Gallagher \et 2002; Turner \& Kraemer 2003; Chartas \et 2004;
Gallagher \et 2005).  The Compton-reflection `hump' is the spectral
manifestation of hard \hbox{X-ray} photons emitted in a corona of hot
electrons and reflected off the relatively colder accretion disk
(e.g., see \S~3.5 of Reynolds \& Nowak 2003 and references
therein). The Compton-reflection `hump' feature lies in the rest-frame
\hbox{$\sim$7--60~keV} energy range, peaking at rest-frame
$\sim$30~keV.  Both the \Ka\ line and Compton reflection spectral
components are expected to be prominent if most of the observed
\hbox{X-ray} flux from the source is reflected or scattered.  No clear
sign of either a neutral (at rest-frame 6.4~keV) or an ionized (at
rest-frame \hbox{6.7--6.97~keV}) \Ka\ emission line, or any sign of a
Compton-reflection component, is apparent in the EPIC spectra of
CSO~755.

To constrain the strengths of \Ka\ lines and the Compton `hump' we used
{\sc xspec} to fit the spectra with two models. The first model
contained a power-law with Galactic and intrinsic absorption, a
Compton-reflection component of \hbox{X-rays} reflected off a neutral
disk (using the {\sc pexrav} model; Magdziarz \& Zdziarski 1995), and
a neutral narrow Gaussian \Ka\ line that had a fixed width of
$\sigma$=0.1~keV and a fixed rest-frame energy of 6.4~keV during the
fit. The second model contained a power-law with Galactic and
intrinsic absorption, a Compton-reflection component of \hbox{X-rays}
reflected off an ionized disk (using the {\sc pexriv} model; Magdziarz
\& Zdziarski 1995), and an ionized narrow Gaussian \Ka\ line that had
a fixed width of $\sigma$=0.1~keV and a fixed rest-frame energy of
6.7~keV during the fit. The photon indices and intrinsic absorption
columns in the two fits were not significantly different than the ones
obtained in \S~\ref{spectral}. We obtained upper limits (90\%
confidence) on the rest-frame equivalent width of a neutral \Ka\ line
of 180~eV, and on the rest-frame equivalent width of an ionized
\Ka\ line of 120~eV.  Similarly, we obtained an upper limit on the
relative Compton-reflection parameter of \hbox{X-rays} reflected off a
neutral disk, $R\leq$0.2. For the {\sc pexriv} model we obtained a
relative Compton-reflection parameter off an ionized disk of
$R$=1.1$^{+1.5}_{-0.8}$. However, an \hbox{$F$-test} indicates that
the existence of this component is not warranted by the data, and that
the marginal improvement in the fit [$\chi^2$(DOF)=99.6(118)],
relative to the fit of \S~\ref{spectral} (Table~\ref{0201_0401_fit}),
is probably due to the ionized iron edge which is included in the {\sc
  pexriv} model; the significance of this edge's existence was already
investigated above. Motivated by recent claims of relativistic broad
absorption lines in the \hbox{X-ray} spectra of some BALQSOs (e.g.,
Chartas \et 2002, 2003), we searched for additional absorption
features in the X-ray spectrum of CSO~755, but our spectrum is
featureless.

\subsection{X-ray `Loudness'}
\label{aox}

Using the SDSS spectrum of CSO~755 we computed the optical-to-X-ray
spectral energy distribution (SED) parameter (or \hbox{X-ray} `loudness'
parameter), \aox, defined~as:

\begin{equation}
\label{eq_aox}
\alpha_{\rm ox}=\frac{\log(f_{\rm 2~keV}/f_{2500\mbox{\rm~\scriptsize\AA}})}
{\log(\nu_{\rm 2~keV}/\nu_{2500\mbox{\rm~\scriptsize\AA}})}
\end{equation}
where $f_{\rm 2~keV}$ and $f_{2500\mbox{\rm~\scriptsize\AA}}$ are the
monochromatic fluxes at rest-frame 2~keV and 2500~\AA, respectively
(e.g., Tananbaum \et 1979). We estimated the monochromatic flux at
rest-frame 2500~\AA\ by extrapolating the monochromatic flux at
rest-frame 1350~\AA\ assuming a continuum of the form $f_{\nu}\propto
\nu^{-\alpha}$ with $\alpha=0.5$ (Vanden Berk \et 2001). This
continuum slope is consistent with the one we measured from the SDSS
spectrum ($\alpha \approx 0.4-0.6$) that shows no indication for
intrinsic reddening.  The monochromatic flux at rest-frame 2~keV was
computed using the power-law normalization at 1~keV of the pn
detector, 4.5$\times$10$^{-5}$~keV~cm$^{-2}$~s$^{-1}$~keV$^{-1}$, and
the spectral parameters of the joint fit of all EPIC detectors in the
first observation (Table~\ref{0201_0401_fit}). We find an observed
(i.e., corrected for Galactic absorption, but {\em not} corrected for
intrinsic absorption) \aox=$-$1.54.  When correcting for intrinsic
absorption we find \aox=$-$1.51, which is consistent with the mean
absorption-corrected value, \aox=$-$1.58, for a non-uniform sample of
seven BALQSOs, for which \aox\ could be measured from high-quality
\hbox{X-ray} observations (Gallagher \et 2002).  Note that the
intrinsic-absorption correction to \aox\ for CSO~755 is relatively
small owing to its small \hbox{X-ray} absorption column density (see
\S~\ref{spectral}). Given the large optical luminosity of CSO~755, the
\aox\ we find is also consistent with \aox=$-$1.73 expected for the
source by the relationship between \aox\ and the monochromatic
luminosity at 2500~\AA\ (Eq.~6 of Strateva \et 2005). Although the
difference between \aox=$-$1.51 and \aox=$-$1.73 corresponds to a
difference of $\sim$3.7 in \hbox{X-ray} flux (Eq.~\ref{eq_aox}), this
is within the scatter around the Strateva \et (2005) relation.  In
fact, CSO~755 was selected for the \xmm\ observations as being one of
the most \hbox{X-ray} luminous quasars, and hence its relatively high
(i.e., less negative) \aox; using our first observation, correcting
for Galactic absorption and for the best-fit neutral intrinsic
absorption we found in \S~\ref{spectral}, we measure $L_{2-10~{\rm
    keV}}$=8.4$\times$10$^{45}$~ergs~s$^{-1}$. This luminosity exceeds
the \hbox{2--10}~keV luminosities of the ten sources in the Shemmer
\et (2005) study, and is exceeded by only one source from the Vignali
\et (2005) study; the sources in those two studies are among the most
\hbox{X-ray} luminous known.

\subsection{X-ray Variability}
\label{variability}

Motivated by the recent discovery of pronounced \hbox{X-ray}
variations in the BALQSO PG~2112$+$059 (Gallagher \et 2004b), we
tested whether CSO~755 exhibits long or short-term \hbox{X-ray} flux
variations.  Brandt \et (1999) report a \hbox{2--10~keV} flux of
1.3$\times$10$^{-13}$~ergs~cm$^{-2}$~s$^{-1}$ for the {\sl BeppoSAX}
observations of the source. For our first observation we find a
\hbox{2--10~keV} flux of
(1.50$^{+0.14}_{-0.16}$)$\times$10$^{-13}$~ergs~cm$^{-2}$~s$^{-1}$ for
the EPIC pn detector (90\% confidence). Similarly, we find a
\hbox{2--10~keV} flux of
(1.58$^{+0.70}_{-0.56}$)$\times$10$^{-13}$~ergs~cm$^{-2}$~s$^{-1}$ for
the EPIC pn detector in our second observation.  These three flux
measurements were made in the \hbox{2--10~keV} observed-frame energy
range, and were not corrected for either Galactic or intrinsic
absorption.  We also compared the total count rates above the
background level for each EPIC detector in each of our two
\xmm\ observations. The results are consistent with the flux
measurements, i.e., $\ltsim$10\% flux variations within a rest-frame
timescale of $\sim$1~d (see Table~\ref{obs_log}). The fluxes we
measure in the two \xmm\ observations are also consistent with the
{\sl BeppoSAX} measurement, given the relative uncertainty between
cross-calibrations of the two observatories (e.g., Kirsch \et 2004),
suggesting that flux variations may be $\ltsim$10\% even on a
$\sim$1~yr rest-frame timescale (although there is scope for
variability in between those two epochs). Using {\sc
  pimms}\footnote{Portable Interactive Multi-Mission Simulator at
  http://heasarc.gsfc.nasa.gov/Tools/w3pimms.html}, we found that the
\hbox{X-ray} flux of CSO~755 obtained from the tentative {\sl ROSAT}
detection is also consistent with the fluxes measured from the {\sl
  BeppoSAX} and \xmm\ observations.  We note that a 10~ks
\xmm\ guaranteed time observation of CSO~755, carried out on
\hbox{2001 July 30}, does not allow a meaningful flux measurement for
variability purposes, since the entire observation is ruined by high
background flaring.  Finally, we searched for rapid ($\sim$1~hr
timescale in the rest frame) variability within our first
\xmm\ observation applying a Kolmogorov-Smirnov test to the photon
arrival times, but none was detected. We also created a light curve
for that observation, binning the photon arrival times in bins of
100~s; the excess variance (e.g., Nandra \et 1997) was consistent with
zero.

\begin{figure*}
\plotone{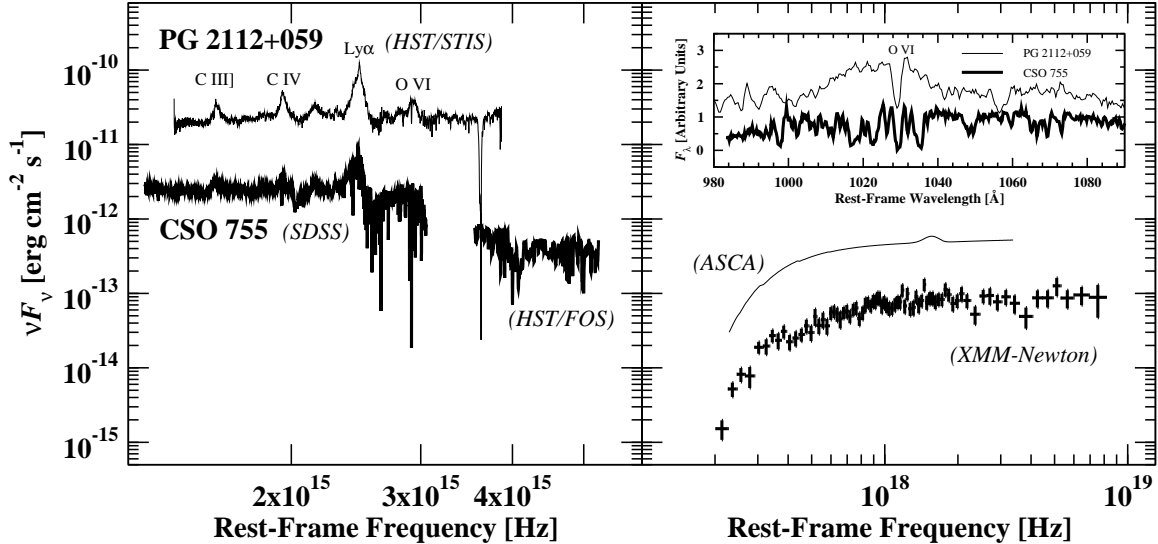}
\caption
{The observed SED of CSO~755 ({\it thick lines}) plotted along with
  the observed SED of PG~2112$+$059 ({\it thin lines}). For clarity,
  only the \hbox{X-ray} observations of the first epoch ({\sl ASCA})
  are shown for PG~2112$+$059 (see Gallagher \et 2004b).  All the SEDs
  are dereddened to remove Galactic absorption. The insert shows the
  \ion{O}{6}~$\lambda$1034 region of the two sources. Note the
  remarkable similarity between the \hbox{UV--X-ray} SEDs of the two
  sources.}
\label{figure_SED}
\end{figure*}

\section{Discussion}
\label{discussion}
\subsection{X-ray Spectrum}
\label{continuum}

Our \xmm\ spectrum of CSO~755 is~among~the three highest
signal-to-noise \hbox{X-ray} spectra of~BALQSOs to date (see Chartas
\et 2002, 2003; Table~\ref{obs_log}). The~photon~index
$\Gamma$=1.83$^{+0.07}_{-0.06}$ and the absorption-corrected
\hbox{optical-to-X-ray} spectral slope \aox=$-$1.51 show that this
source has an \hbox{X-ray} continuum of a `typical' radio-quiet quasar
(e.g., Vignali \et 2005). Our results are therefore in line with the
conclusion that the \hbox{X-ray} continua of BALQSOs are not
intrinsically different from those of the majority of the quasar
population.

We find that the \hbox{X-ray} spectrum of CSO~755 is moderately
absorbed, with hints of a complex absorption pattern at rest-frame
energies $\ltsim1.5$~keV. The best-fit neutral intrinsic column
density is \nh$\sim$1.2$\times$10$^{22}$~cm$^{-2}$, and our data do
not allow us to rule out the possibility that there is some absorption
by ionized gas, with column densities of
$\approx$10$^{23}$~cm$^{-2}$. Even though we cannot rule out the
possibility that there are other \hbox{X-ray} components in the source
which are much more absorbed, the \aox=$-$1.51 we obtain is already
quite flat (i.e., more \hbox{X-ray} bright) relative to the
\aox\ found in radio-quiet quasars with similar
luminosities. Any additional absorption would have required the
intrinsic \aox\ to be even flatter, which is unlikely.  Even
though the redshift of CSO~755 gives us a good opportunity to measure
the high-energy iron features with \xmm\ to respectable accuracy, we
did not detect any features, such as the relativistic broad Fe
absorption lines perhaps observed in the BALQSOs APM~08279$+$5255
(Chartas \et 2002) and PG~1115$+$080 (Chartas \et 2003).  The only
\hbox{X-ray} manifestation of the UV absorber in CSO~755 therefore
lies in the relatively low intrinsic neutral column we detect at low
energies. Our results bolster the idea that BALQSOs display a range of
intrinsic columns ($\approx$10$^{22}$--10$^{24}$~cm$^{-2}$), and a
broad range of \hbox{X-ray} spectral features.

The absence of prominent iron lines or a Compton-reflection `hump' in
the spectrum of CSO~755 indicates that \hbox{X-ray} reflection from
the disk into our line-of-sight is weak or absent.  The idea that most
of the \hbox{X-rays} are directly observed is also indicated by the
rather flat \aox\ we find. Since most of the \hbox{X-rays} are
directly observed, and are not scattered or reflected, one might have
expected the source to exhibit some level of \hbox{X-ray} flux
variations. However, we do not detect significant variability on
either short ($\ltsim$1~d) or long ($\sim$1~yr) timescales. This may
be an indicator of a large black-hole mass, \mbh$\gtsim$10$^{9}$\msun,
in the source (e.g., O'Neill \et 2005).  In fact, assuming
Eddington-limited accretion, we estimate a black hole mass of
\mbh$\gtsim$5$\times$10$^9$\msun\ from the optical luminosity of the
source.

\subsection{UV--Optical Polarization and \hbox{X-ray} Properties}
\label{polarization}

Based upon the relatively high \hbox{UV--optical} polarization level
of CSO~755 ($P_{V}\simeq$3.5\%), it might have been possible for most
of the observed \hbox{X-ray} flux from the source to be scattered or
reflected. The fact that the bulk of the \hbox{X-ray} emission from
CSO~755 appears to be directly observed, however, calls for an
investigation of the relationship between \hbox{UV--optical}
polarization and \hbox{X-ray} properties in a larger sample of
BALQSOs. We have compiled a non-uniform sample of 46 BALQSOs from the
literature (including the CSO~755 data presented in this paper) which
have published \hbox{X-ray} properties (Gallagher \et 2002; Grupe \et
2003; Gallagher \et 2004a; Gallagher \et in prep.)  and corresponding
\hbox{UV--optical} polarization measurements (Hutsem{\' e}kers \et
1998; Ogle \et 1999; Schmidt \& Hines 1999; Hutsem{\' e}kers \& Lamy
2000, 2001). We have not found significant correlations between the
broad-band \hbox{UV--optical} polarization level and any of the
following \hbox{X-ray} properties: \aox, hardness ratio, photon index,
neutral intrinsic absorption, flux, and luminosity. In particular, the
suggestion that highly polarized BALQSOs are also the \hbox{X-ray}
brightest (or more \hbox{X-ray} luminous) is not supported by our
analysis (Gallagher \et 1999). We caution, however, that the
\hbox{UV--optical} polarization data are not homogeneous, and are
subject to considerable measurement uncertainties.

\subsection{Comparison with the BALQSO PG~2112$+$059}
\label{pg2112}

Some of the \hbox{UV--X-ray} properties of CSO~755 are similar to those of
the well-studied BALQSO PG~2112$+$059 (Gallagher \et 2001, 2004b). For
example, both sources display shallow BAL troughs in the UV, and their
neutral intrinsic column densities, \nh$\sim$10$^{22}$~cm$^{-2}$, are
among the lowest columns observed in BALQSOs. It is therefore possible
that shallow BAL troughs are associated with relatively mild
\hbox{X-ray} absorption; we are not aware of any counterexamples.  To
test this possibility, high-quality \hbox{X-ray} observations of more
BALQSOs with shallow troughs are required.

In Fig.~\ref{figure_SED} we plot the observed SED of CSO~755 along
with that of PG~2112$+$059. The SEDs have been corrected for Galactic
absorption. One can see that the general shapes~of~the UV spectra of
the two sources are quite similar to~one~another, and that a
remarkable similarity is apparent between the {\sl ASCA} spectrum of
PG~2112$+$059 and our \xmm\ spectrum of CSO~755. In fact, scaling down
the {\sl ASCA} model for PG~2112$+$059 by a factor of $\sim$6 in flux
makes it almost identical to our \xmm\ spectrum of CSO~755.~The~UV
spectrum of CSO~755 is scaled down by an almost~constant factor of
$\sim$10 in flux relative to the UV spectrum of PG~2112$+$059. The
difference between the \hbox{X-ray} and UV scaling factors is also
consistent with the difference in \aox\ of the two sources (see
\S~\ref{aox}; Gallagher \et 2004b). The main differences between the
two sources lie in the details of their UV spectra, which are apparent
in the insert to Fig.~\ref{figure_SED}, and the fact that in contrast
with CSO~755, PG~2112$+$059 is a low-polarization BALQSO, though
polarization data for this source are not available in the literature
covering the same rest-frame wavelengths as for CSO~755. In addition,
as opposed to PG~2112$+$059, CSO~755 did not show any \hbox{X-ray}
variations. \\ \\

\section{Summary}
\label{summary}
We have presented new \xmm\ observations of CSO~755 that provide one
of the best BALQSO \hbox{X-ray} spectra.  Our main results can be
summarized as follows:
\begin{enumerate}
\item{The power-law \hbox{X-ray} photon index we found,
  $\Gamma$=1.83$^{+0.07}_{-0.06}$, is consistent with photon indices
  observed for `typical' radio-quiet quasars.}
\item{The intrinsic \aox=$-$1.51 we found is rather flat (\hbox{X-ray}
  bright) but is consistent with the \aox\ distribution observed for
  radio-quiet quasars of similar luminosity.}
\item{By fitting the spectrum with a neutral-absorption model, we
  found a column density of \nh$\sim$1.2$\times$10$^{22}$~cm$^{-2}$,
  which is among the lowest \hbox{X-ray} columns measured for
  BALQSOs.}
\item{We did not detect, with high significance, any other emission or
  absorption features in the X-ray spectrum. The lack of signatures of
  either an \Ka\ line or a Compton-reflection component suggests that
  most of the \hbox{X-rays} from the source are directly observed
  rather than being scattered or reflected, as might have been
  expected given the source's high level of \hbox{UV--optical}
  polarization; this is also supported by the relatively flat
  intrinsic \aox\ we measured.}
\item{We found that the \hbox{UV--optical} continuum polarization
  level of BALQSOs is not correlated with any of their \hbox{X-ray}
  properties, based on a sample of 46 BALQSOs from the literature,
  including CSO~755.}
\item{A lack of significant short- and long-term \hbox{X-ray} flux
  variations in CSO~755 may be attributed to a large black-hole mass
  in the source.}
\item{We note that both CSO~755 and another luminous BALQSO,
  PG~2112$+$059, display both shallow \ion{C}{4} BAL troughs and
  moderate \hbox{X-ray} absorption, suggesting a possible relationship
  between these two properties.}
\end{enumerate}

\acknowledgments

This work is based on observations obtained with \xmm, an ESA science
mission with instruments and contributions directly funded by ESA
Member States and the USA (NASA). We gratefully acknowledge the
financial support of NASA grant \hbox{NAG5--9932} (OS, WNB), NASA LTSA
grant \hbox{NAG5--13035} (WNB), and MIUR COFIN grant 03-02-23 (AC,
CV).  Support for SCG was provided by NASA through the {\em Spitzer}
Fellowship Program, under award 1256317. We are grateful to Shai Kaspi
and Doron Chelouche for useful discussions. We also appreciate the
constructive comments made by the referee, Mike Brotherton, that
helped to improve the manuscript.

\end{document}